\documentclass{iopjournal}
\usepackage{textcomp}
\usepackage{stmaryrd}
\usepackage{amsmath}
\usepackage{amssymb}
\usepackage{caption}
\usepackage{subcaption}
\usepackage{color}
\usepackage{bm}
\usepackage{threeparttable}
\usepackage{booktabs}
\usepackage{algorithmic}
\usepackage{cite}
\usepackage{mathtools}
\usepackage{textcomp}
\usepackage{stfloats}
\usepackage{url}
\usepackage{verbatim}
\usepackage{eqnarray}
\usepackage{graphicx}
\usepackage{balance}
\usepackage{graphics}
\usepackage[dvipsnames]{xcolor}
\usepackage{dcolumn}
\usepackage[utf8]{inputenc}
\usepackage{textcomp}
\usepackage{footnote}
\usepackage{tablefootnote}
\usepackage{hologo}
\usepackage{booktabs}
\usepackage{epstopdf}
\usepackage{url}
\usepackage{csquotes}
\usepackage[british]{babel}
\DeclareUnicodeCharacter{2212}{-}
\DeclareUnicodeCharacter{03B4}{$\delta$}
\usepackage{siunitx} 
\usepackage[switch]{lineno}
\usepackage{lmodern} 
\usepackage{fix-cm}
\usepackage{multirow}
\usepackage{tikz}
\usepackage{soul}

\def\BibTeX{{\rm B\kern-.05em{\sc i\kern-.025em b}\kern-.08em
    T\kern-.1667em\lower.7ex\hbox{E}\kern-.125emX}}
\newcommand{\R}{\mathbb{R}}
\newcommand{\curl}{\mathop{\nabla\times}}
\newcommand{\divg}{\mathop{\nabla\cdot}}
\newcommand{\dd}{\,\mathrm{d}}
\newcommand{\Jc}{J_{\mathrm c}}
\newcommand{\Ec}{E_{\mathrm c}}

\newcommand{\Hcurl}{H(\mathrm{curl};\Omega)}
\newcommand{\Hcurlz}{H_0(\mathrm{curl};\Omega)}
\begin{document}

\articletype{Letter} 

\title{A Cloud-Accessible Open-Source Framework for the Electromagnetic Modelling of Applied Superconductors}

\author{Yusen Guo$^{1}$\orcid{0000-0003-2354-0395}, Alberto Paganini$^{1*}$\orcid{0000-0003-3309-7657} and Harold S. Ruiz$^{2*}$\orcid{0000-0002-6100-1918}}

\affil{$^{1}$School of Computing and Mathematical Sciences, University of Leicester, Leicester LE1 7RH, UK.}

\affil{$^{2}$School of Engineering and Space Park Leicester, University of Leicester, Leicester LE1 7RH, UK.}

\affil{$^{*}$ email \href{mailto:a.paganini@leicester.ac.uk}{a.paganini@leicester.ac.uk}, \href{mailto:dr.harold.ruiz@leicester.ac.uk}{dr.harold.ruiz@leicester.ac.uk}}


\begin{abstract}
We present the \emph{H-cloud} formalism, a cloud-accessible and open-source finite-element framework for electromagnetic modelling of applied superconductors. The proposed method expresses the nonlinear electromagnetic response of type-II superconductors in a curl-conforming discretisation based on N\'ed\'elec finite elements, where the tangential applied-field boundary condition, nonlinear \(E\)--\(J\) power law, and fully implicit time-discrete residual are stated explicitly at the variational level, all within a scripted Python finite-element workflow. The weak form is used as the basis for forward simulation and for extension to adjoint differentiation and PDE-constrained optimisation, while keeping the governing assumptions, boundary conditions, and solver structure fully visible to the user. The implementation is realised in Firedrake with UFL and PETSc-backed nonlinear solvers, allowing the identical script to run on local machines and in browser-accessible environments such as Google Colab without reformulating the problem. The method is verified on the canonical magnetisation benchmark of a cylindrical superconductor under Bean-like penetration conditions and then benchmarked against an independently constructed COMSOL model for a practical high temperature superconducting Bi2212 wire. Across matched mesh studies, the open-source workflow reproduces the commercial-reference magnetisation loops to within approximately \(1\%\) , with relative peak errors below \(1.5\%\), while cloud execution preserves the same numerical solution at rather modest additional runtime considering the use of (freely available) reduced hardware resources. The proposed framework provides a rigorous, reproducible, and portable route for superconducting simulation, benchmarking, and future optimisation-led modelling of applied and functional superconductors, shareable and executable into open cloud environments.
\end{abstract}

\section{Introduction}\label{Sec.1}\vspace*{.2cm}

Electromagnetic modelling is indispensable for contemporary superconducting technology. Whether the target device is a magnetic cloak~\cite{Gomory2012,Ruiz2025-SciAdv}, an HTS cable or coil~\cite{Driscoll_21}, a rotating machine~\cite{Coombs2024NatRev}, a fault-current limiter~\cite{Ruiz2015-IEEETAS-SFCL}, or another field-control component such as an HTS dynamo or flux pump~\cite{Brooks_2020,Mataira2019}, the design problem is ultimately controlled by transient flux penetration~\cite{Ruiz2018-SUST-DCwires,Ruiz2009-PRB-GCST}, shielding-current redistribution~\cite{Ruiz2014-APL}, hysteretic magnetization~\cite{Ruiz2012-APL,Ruiz2018-SUST-DCwires}, and nonlinear $E$--$J$ electrodynamics~\cite{Ruiz2025-PMS}. Reliable numerical studies therefore require repeated solutions of magneto-quasi-steady Maxwell problems over changing geometries, loading histories, and constitutive parameters~\cite{GrilliIEEE21,Wimbush_SuST22}. Since analytical solutions exist only for idealised benchmark geometries and limiting critical-state conditions~\cite{Bean1962,Ruiz2009-PRB-GCST}, finite-element and variational numerical models remain essential for evaluating local current density~\cite{Ruiz2025-PMS}, magnetic-field penetration, magnetization, and AC-loss behaviour in realistic superconducting components~\cite{ Driscoll_21,Coombs2024NatRev}.

Among the formulations used for applied superconductivity, the $H$-formulation is one of the most established because it represents the magnetic field directly and combines naturally with the power-law $E$--$J$ relation commonly used for flux creep~\cite{ShenBoyang2020OoHA,GrilliIEEE21}. Yet, it is not the only viable framework. Depending on the conductor topology and modelling target, especially when either thin-film approximations, transport-current constraints, reduced non-superconducting domains, or large superconducting tape assemblies dominate the computational requirements, researchers also use vector-potential approaches such as the $T$--$A$~\cite{GrilliIEEE21,Benkel2020,Zhang2017}, $A$--$V$ and $A$--$\phi$ formulations~\cite{RuizAlonso2004,Lahtinen2012}, mixed $H$--$A$ and $H$--$\phi$ strategies~\cite{Dular2021MixedFE,Arsenault2021ReducedHphi}, $J$-based integral and variational methods~\cite{Brambilla2009,Pardo_JoC_2017}, and minimum-electromagnetic-entropy-production-type approaches~\cite{Badia_SUST_2012,Ruiz2009-PRB-GCST}, all aiming to reduce computational cost, enforce transport-current constraints, avoid unnecessary air-domain discretisation, or treat large-scale superconducting assemblies and three-dimensional geometries within affordable computational architectures. The appropriate choice depends on the scale of the device, the dimensional reduction that can be justified physically, the presence of materials with different constitutive laws, the level of control required over the numerical solver, and the required balance between accuracy, transparency, and computational cost.

Compared with the extensive literature on alternative superconducting formulations, less attention has been given to formulations in which the full superconducting model can be written transparently from the weak form upward and, how to make it immediately portable to a cloud-executable workflow. This point matters scientifically as much as practically. Commercial multiphysics platforms can provide efficient forward solutions and have been instrumental in the dissemination of superconducting finite-element modelling~\cite{ShenBoyang2020OoHA,GrilliIEEE21}, but they are less well suited when the goal is to expose the variational structure, control the solver configuration in detail, link forward simulation to adjoint differentiation~\cite{Pyadjoint2019}, and distribute the exact same calculation to collaborators without relying on a local licensed environment, thereby serving as a reusable computational reference rather than as a solver-specific recipe. Conversely, open-source finite-element implementations can offer that level of transparency, but can still be difficult to install and reproduce across machines. This is particularly limiting for benchmark calculations, where the applied-field boundary condition, nonlinear resistivity law, mesh, time step, and solver tolerances should remain visible and modifiable.

The present paper introduces this viewpoint through what we call the \emph{H-cloud formalism}, recasting the well-established magnetization benchmark for cylindrical superconducting wires under Bean's model~\cite{Bean1962,Bean1964,Karmakar2003,Ruiz2012-PhDThesis}, yet through a cloud-accessible open-source workflow built from the variational residual upward from a curl-conforming discretisation of the $H$-formulation. The implementation is written in a Python-based finite-element environment, here using Firedrake~\cite{FiredrakeUserManual}, where weak forms are expressed in the Unified Form Language (UFL) and solved through PETSc-backed nonlinear algorithms~\cite{Rathgeber2016,PETScWebPage}. The same script can be executed locally or inside browser-accessible notebook environments such as Google Colab~\cite{GoogleColab}, which makes the computational pathway easier to share and reproduce. The cloud aspect is therefore not a wrapper added after the mathematics; it is a consequence of how the mathematics has been written, with the corresponding executable workflow being provided in the \emph{Supplementary Material}.

The contribution is threefold. First, we state the superconducting $H$-formulation problem in a self-contained weak form posed in a curl-conforming function space $\Hcurl$, with the boundary condition applied through the tangential trace of the chosen space. Second, we show how the weak form can be advanced as a fully discrete forward solver of variational nature, embedded in a partial differential equation (PDE)-constrained optimisation framework. Third, we verify the implementation on the canonical Bean-cylinder benchmark and compare runtime and accuracy against an independently constructed COMSOL model, demonstrating that the same workflow runs locally, achieving equitable performance levels, while remaining executable in a modest cloud notebook environment.

\section{Mathematical and Computational Framework}\label{Sec.2}\vspace*{.2cm}

The proposed \emph{H-cloud} formalism and computational method is built in two stages. First, the magneto-quasi-steady $H$-formulation is written in weak form and discretised using a curl-conforming finite-element space, where the induced current density follows from the curl of the computed field, while the applied magnetic excitation is imposed through the tangential trace. Second, the weak problem is turned into a fully discrete nonlinear residual using backward-Euler time stepping. This residual retains a direct link to the mathematical formulation and is the same object implemented in the cloud-executable finite-element workflow. In fact, because this residual is the primary computational object, it can be passed unchanged to nonlinear solvers, post-processing routines, and adjoint differentiation tools, keeping the physical assumptions transparent while making the computational implementation directly reproducible, as it is shown below. 


\subsection{Curl-conforming $H$-formulation in the magneto-quasi-steady regime}\vspace*{.2cm}

Let $\Omega\subset\R^{d}$, with $d=2$ or $3$, denote the computational domain. We partition it into a superconducting region $\Omega_{\mathrm{sc}}$ and a non-superconducting region $\Omega_{\mathrm{nsc}}$, and assume that $\Omega$ is a Lipschitz domain with boundary $\partial\Omega$, i.e., locally representable as the graph of a Lipschitz-continuous function. Under the magneto-quasi-steady (MQS) approximation~\cite{Ruiz2009-PRB-GCST}, the displacement current is neglected and the governing equations reduce to Faraday's law and Ampère's law in the standard form for applied superconductors~\cite{Ruiz2025-PMS}, 
\begin{eqnarray}
     \partial_t \mathbf{B} + \curl \mathbf{E} = 0 \, ,\label{eq:faraday}\\
    \mathbf{J} = \curl \mathbf{H} \, ,\label{eq:ampere}   
\end{eqnarray}
where $\mathbf{H}$ is the magnetic field intensity, $\mathbf{B}$ is the magnetic flux density, $\mathbf{E}$ is the electric field, and $\mathbf{J}$ is the current density. To close the electromagnetic problem, we use the magnetic constitutive relation,
\begin{equation}
    \mathbf{B} = \mu \mathbf{H} \, ,
    \label{eq:constitutiveB}
\end{equation}
where $\mu=\mu_0\mu_r$, $\mu_0$ is the permeability of free space, and $\mu_r$ is the relative magnetic permeability of the domain under consideration. In the present benchmark we assume linear isotropic magnetic response and take $\mu_r=1$ throughout, allowing a direct comparison with the analytical solution where the superconducting domain is treated as non-magnetic unless otherwise stated. 

To model the superconducting electrodynamics we use the standard power-law $E$--$J$ relation in $\Omega_{\mathrm{sc}}$, together with a linear resistive law in the non-superconducting region. Writing $\mathbf{E}=\rho(\mathbf{J})\mathbf{J}$ and combining this with \eqref{eq:faraday}--\eqref{eq:constitutiveB} yields
\begin{equation}
    \partial_t(\mu \mathbf{H})+\curl\left(\rho(\mathbf{J})\curl\mathbf{H}\right)=0
    \quad\text{in }\Omega \, ,
    \label{eq:Hform}
\end{equation}
where
\begin{equation}
    \rho(\mathbf{J})=
    \begin{cases}
    \dfrac{\Ec}{\Jc}\left(\dfrac{|\mathbf{J}|}{\Jc}\right)^{n-1}, & \mathbf{r}\in\Omega_{\mathrm{sc}},\\[1.2ex]
    \rho_{\mathrm{nsc}}, & \mathbf{r}\in\Omega_{\mathrm{nsc}}\, .
    \end{cases}
    \label{eq:rho_piecewise}
\end{equation}
Here $\Ec$ is the critical electric-field criterion, $\Jc$ is the critical current density, $n$ is the power-law exponent, $\mathbf{r}$ denotes the element position, and $\rho_{\mathrm{nsc}}$ is the electrical resistivity assigned to the non-superconducting domain.

The magnetic flux conservation condition $\divg\mathbf{B}=0$ is consistent with Faraday's law because the divergence of a curl vanishes, provided the initial magnetic flux density at a time $t_{0}$ is divergence-free, and the MQS approximation remains valid for a time $t>t_{0}$. This can be easily achieved by assuming $\mathbf{H}(t_{0})=\bm 0$, i.e., $\mathbf{B}(t_{0})=\bm 0$ by \eqref{eq:constitutiveB}, and a sufficiently small time-stepping to preserve the variational principles of the MQS regime. Therefore, special attention must be taken on preserving the curl-conforming structure that governs the tangential continuity of the field. In the present weak $H(\mathrm{curl})$ setting, this condition is not introduced as a separate scalar equation; instead, the finite-element space is chosen to preserve the curl structure of the continuous problem in a seamless way.

For the sake of ease, in the computational benchmark considered in this paper the superconducting domain is assumed to be an infinitely long cylinder with finite radius $R$, which is immersed within a transverse magnetic field $\mathbf{H}=(H_x,H_y)$ in the $xy$ plane, while the induced current density results out-of-plane with its magnitude defined by,
\begin{equation}
    J_z = \partial_x H_y-\partial_y H_x \, .
    \label{eq:2dcurl}
\end{equation}
The $E$--$J$ power law relation is then applied to $J_z$, and the external excitation is imposed through a time-dependent tangential boundary trace corresponding to a background field $\mathbf{H}_{\mathrm{bg}}(t)$, which can be defined by the AC function,
\begin{equation}
    \mathbf{H}_{\mathrm{bg}}(t)
    =
    \mu_0^{-1}B_a\sin(\omega t)\,\hat{\mathbf{J}} \, ,
    \label{eq:bc}
\end{equation}
where $B_a$ is the applied-field amplitude and $\omega=2\pi f$ its angular frequency.

To derive a finite element discretization of \eqref{eq:Hform}, we introduce the Hilbert space $\Hcurl$, such that
\begin{equation} \label{eq:Hcurl}
    \Hcurl=\left\{\bm v\in [L^2(\Omega)]^2:\curl\bm v\in L^2(\Omega)\right\}\, ,
\end{equation}
where \(L^2(\Omega)\) denotes the space of square-integrable functions on
\(\Omega\), and \(\bm v\) is a vector field in this space. We also define the homogeneous tangential-trace subspace,
\begin{equation} \label{eq:H0curl}
    \Hcurlz=\left\{\bm v\in\Hcurl:\bm v\times\bm n=0\text{ on }\partial\Omega\right\}\, ,
\end{equation}
where $\bm n$ denotes the outward unit normal on the boundary $\partial\Omega$.
Note that, in the $\Hcurl$ setting, the background field enters as a Dirichlet boundary condition through the tangential trace operator $\gamma_\tau$, with 
\begin{equation}
\gamma_\tau :
H(\mathrm{curl};\Omega)
\longrightarrow
H^{-1/2}(\mathrm{div};\partial\Omega),
\quad
\gamma_\tau \bm{v}:=\bm{v}\times \bm n \, .
\end{equation}
The boundary condition is then prescribed as
\begin{equation}
    \gamma_\tau\mathbf{H}(t)
    =
    \mathbf{H}_\tau(t)
    :=
    \gamma_\tau\mathbf{H}_{\mathrm{bg}}(t)
    \quad\text{on }\partial\Omega \, .
    \label{eq:tangential_bc}
\end{equation}
This makes the applied field fully explicit at the variational level, which will be important in the benchmark analysis of finite-domain truncation.

Then, the weak form of the problem, i.e., mathematically restating the governing partial differential equations in an integral format, results by multiplying \eqref{eq:Hform} by a test function $\bm\psi\in\Hcurl$. Then, after integration by parts, the resulting weak formulation reads: 

Find $\mathbf{H}(t_{n})\in\Hcurl$ satisfying \eqref{eq:tangential_bc} such that,
\begin{equation}
    \left\langle
    \partial_t(\mu\mathbf{H}),\bm\psi
    \right\rangle_\Omega
    +
    \left\langle
    \rho(J)\curl\mathbf{H},\curl\bm\psi
    \right\rangle_\Omega
    =
    0
    \quad \forall \quad \bm\psi\in\Hcurl.
    \label{eq:weak}
\end{equation}
The boundary contribution \(\langle \rho(J)\left(\curl\mathbf{H}\right)\times \bm n,  \bm \psi \rangle_{\partial\Omega}\) arising from integration by parts, vanishes because the test functions have homogeneous tangential trace. Also, \eqref{eq:weak} results from further assuming that the quantity $\rho\, (\nabla \times \mathbf{H}) \times \mathbf{n}$ is continuous across all interfaces. In this regard, while the corresponding continuity of the tangential electric field is imposed weakly through the posed variational principles, by enforcing $\mathbf{H}(t_{n})$ in $\Hcurl$ using N\'ed\'elec elements, the degrees of freedom are map to the line integrals of $\mathbf{H}$ along the element edges, ergo ensuring tangential continuity at the interfaces. This means, that at any discrete time step $t_{n}$, the tangential continuity of $\mathbf{H}(t_{n})$ guarantees that no unphysical surface currents are artificially generated at the elements between interfaces.


\subsection{Conforming finite-element discretisation and cloud-executable optimisation form}\vspace*{.2cm}

We begin by introducing the finite elements used in this formulation. Traditional Lagrange elements, whose degrees of freedom are associated with mesh nodes, are not the most natural choice for discretising Maxwell-type problems in $\Hcurl$. Although they are technically $\Hcurl$-conforming, they may produce spurious solutions because they do not belong to a de Rham complex~\cite{Ar18}. In contrast, N\'ed\'elec elements, commonly referred to as edge elements, are specifically designed for $\Hcurl$ spaces. Their construction ensures continuity of the tangential component of the vector field across element boundaries, a property that is essential for the accurate discretisation of electromagnetic problems.

Consequently, let $\mathcal T_h$ be a conforming mesh of $\Omega$ and let $\mathcal{N}_{h} \subset\Hcurl$ denote the corresponding N\'ed\'elec finite element subspace. Then, by defining $\Delta t>0$ as the time step, we can set $t_k=k\Delta t$, and let $\mathbf{H}_h^k\in \mathcal{N}_{h}$ denote the finite-element approximation to $\mathbf{H}(t_k)$. Thus, for fully implicit backward Euler time stepping, given $\mathbf{H}_h^k$, we find $\mathbf{H}_h^{k+1}\in \mathcal{N}_{h}$ by imposing \eqref{eq:tangential_bc} at $t_{k+1}$ and solving the nonlinear variational problem:
\begin{equation}
\begin{aligned}
    F(\mathbf{H}_h^{k+1};\bm\psi_h) & := \left\langle\mu\frac{\mathbf{H}_h^{k+1}-\mathbf{H}_h^k}{\Delta t},\bm\psi_h\right\rangle_\Omega 
    +\left\langle\rho(J_z^{k+1})\curl\mathbf{H}_h^{k+1},\curl\bm\psi_h\right\rangle_\Omega=0 \, ,
\end{aligned}
\label{eq:be}
\end{equation}
for all admissible $\bm\psi_h\in \mathcal{N}_{h}^0$, where $\mathcal{N}_{h}^0$ is the discrete test space with homogeneous tangential boundary condition within $\Hcurl$. 

The residual in \eqref{eq:be} is the central computational object of the \emph{H-cloud} formalism: it is simultaneously the weak integro-differential statement of the superconducting problem, the nonlinear algebraic system advanced at each time step, and the equality constraint used when the same solver is embedded into a PDE-constrained optimisation problem. In practice, rather than beginning from software-specific menus or PDE templates, we express this residual directly in a standard UFL format, which is a high-level language used to define and express mathematical PDEs in a notation that closely resembles standard pen-and-paper mathematics. Then, at each time step, the fully discrete nonlinear algebraic system is solved by the Newton--Kantorovich algorithm with line search using PETSc \cite{PETScWebPage}, and Firedrake as the finite-element engine \cite{FiredrakeUserManual,Rathgeber2016}. Because the workflow is scripted rather than GUI driven, the mesh, timestep, constitutive law, solver tolerances, and diagnostics can all be modified from within the same Python notebook or local script (\emph{Supplementary Material}), without altering the underlying mathematics. Consequently, the same mathematical model can be executed locally or in a cloud notebook without changing the formulation.

Furthermore, a rather significant advantage of this method is that the same forward variational formulation can be embedded in a PDE-constrained control optimisation problem by treating the $H$-formulation residual as an equality constraint with a well-defined control variable $\eta$. For instance, take as  reference our recently proposed method for designing functional magnetic cloaks at~\cite{Ruiz2025-SciAdv}, where a spatially varying material parameter such as the relative magnetic permeability of a soft-ferromagnetic material was introduced as the control variable. Thus, more broadly, the forward problem \eqref{eq:weak} may be written as: find \(\mathbf{H}=\mathbf{H}(\eta)\subset H(\mathrm{curl};\Omega)\) such that the state equation can be defined by,
\begin{equation}
    F(\mathbf{H},\eta; \bm \psi)=0
    \quad \forall \quad \bm \psi \in \Hcurlz ,
    \label{eq:state_residual}
\end{equation}
where \(\Hcurlz\) is the corresponding test space with homogeneous tangential boundary condition, and \(\mathbf{H} \) is the linked magnetic field state. Then, for a control variable $\eta$, one may define
\begin{equation}
    \mathcal J(\mathbf{H},\eta)=
    \frac{1}{2}\sum_{k=1}^{N_t}\Delta t
    \int_{\Omega_{\mathrm{obs}}}|\mathbf{H}^k-\mathbf{H}_{\mathrm{target}}^k|^2\dd x
    +\mathcal R(\eta)\, ,
    \label{eq:objective}
\end{equation}
where $\Omega_{\mathrm{obs}}$ denotes the observation region, $\mathbf{H}_{\mathrm{target}}^k$ is the target field at time step $k$, and $N_{t}$ is the total number of time steps included in the minimisation. The first term quantifies the field-matching error in $\Omega_{\mathrm{obs}}$, whereas the second term, $\mathcal R(\eta)$, introduces a smoothed total-variation regularisation of the control variable $\eta$. In the language of variational control theory, this term acts as an additional penalty in the cost functional, suppressing non-physical oscillations and discouraging strongly mesh-dependent control patterns. Its inclusion is therefore important for the mathematical stability of the optimisation problem and for the engineering interpretability of the resulting design. Without such regularisation, the inverse problem may favour highly anisotropic or numerically irregular solutions that minimise the objective but are not robust, manufacturable, or physically meaningful.

Consequently, the PDE-constrained optimisation problem can be posed in general terms as, 
\begin{equation}
    \min_{\eta\ | \ R(\eta)}
    \widehat{\mathcal{J}}(\eta)
    :=
    \mathcal{J}(\mathbf{H}(\eta),\eta)
    \quad
    \mathrm{subject\ to}
    \quad
    F(\mathbf{H}(\eta),\eta;\bm \psi)=0
    \quad \forall \quad \bm \psi\in \Hcurlz\, .
    \label{eq:pde_constrained_problem}
\end{equation}
Here, it is to be noted that a significant computational advantage emerges, as the gradient of the so-called reduced functional \(\widehat{\mathcal{J}}\) can be obtained without explicitly differentiating the state
\(\mathbf{H}(\eta)\) with respect to every degree of freedom of \(\eta\). Furthermore, for a finite control perturbation \(\delta\eta\), let
\(\delta\mathbf H:=\mathbf H'[\eta]\delta\eta\in \Hcurlz\) denote the corresponding
state sensitivity. Differentiating the state residual \eqref{eq:state_residual} gives the tangent-linear equation, 
\begin{equation}
    F_{\mathbf{H}}(\mathbf{H},\eta;\delta\mathbf{H},\bm\psi)
    +
    F_{\eta}(\mathbf{H},\eta;\delta \eta,\bm\psi)
    =
    0
    \quad \forall \quad \bm\psi\in \Hcurlz\, .
    \label{eq:tangent_linear}
\end{equation}
Here, a direct sensitivity approach would require solving \eqref{eq:tangent_linear} once for each independent component of \(\eta\), which is computationally infeasible for high-dimensional finite-element controls. Nonetheless, an adjoint variable \(\bm\lambda\in \Hcurlz\) is introduced and computed from
\begin{equation}
    F_{\mathbf{H}}(\mathbf{H},\eta;\bm\phi,\bm\lambda)
    =
    -\mathcal{J}_{\mathbf{H}}(\mathbf{H},\eta;\bm\phi)
    \quad \forall \quad \bm\phi\in \Hcurlz \, . 
    \label{eq:adjoint_equation}
\end{equation}
Using the chain rule, the reduced derivative is
\begin{equation}
    \widehat{\mathcal{J}}'(\eta;\delta\eta)
    =
    \mathcal{J}_{\mathbf{H}}(\mathbf{H},\eta;\delta\mathbf{H})
    +
    \mathcal{J}_{\eta}(\mathbf{H},\eta;\delta\eta).
    \label{eq:reduced_derivative_chain_rule}
\end{equation}
Taking \(\bm\phi=\delta\mathbf{H}\) in \eqref{eq:adjoint_equation} and
\(\bm\psi=\bm\lambda\) in \eqref{eq:tangent_linear} eliminates the state
sensitivity, giving
\begin{equation}
    \widehat{\mathcal{J}}'(\eta;\delta\eta)
    =
    \mathcal{J}_{\eta}(\mathbf{H},\eta;\delta\eta)
    +
    F_{\eta}(\mathbf{H},\eta;\delta\eta,\bm\lambda).
    \label{eq:reduced_gradient}
\end{equation}
Thus, one forward-solve and one adjoint-solve provide the derivative with respect to all control degrees of freedom, making this model particularly suitable for high-dimensional material and shape design problems as demonstrated at~\cite{Ruiz2025-SciAdv}. In this implementation, this can be automated through Pyadjoint which records the forward solve and automatically constructs the corresponding adjoint equation \cite{Pyadjoint2019}. The computed derivative may then be passed to a gradient-based optimiser such as the Broyden--Fletcher--Goldfarb--Shanno (BFGS) algorithm \cite{SciPy2020} within the Firedrake environment.

To provide a clear and physically interpretable validation, below we focus on the forward solver and assesses it against a well-established superconducting benchmark. Extensions to multi-objective inverse-design functionals, of the type explored in~\cite{Ruiz2025-SciAdv}, are therefore deferred to future studies.

\section{Results}\label{Sec.3}\vspace*{.2cm}

We now assess the framework from two complementary viewpoints. First, we verify that the formulation reproduces the classical full-penetration scale for an infinitely long superconducting cylinder in transverse applied field, which remains a standard validation test because it combines a nontrivial current penetration pattern with a clear analytical reference. Second, we compare the resulting implementation against an independently constructed COMSOL model and examine the cost of running the same open-source workflow locally and in the cloud.


\subsection{Magnetisation benchmark}\vspace*{.2cm}

For the cylindrical benchmark, when the AC magnetic field is applied along a single direction, say $H_y$, the principal diagnostics is the magnetization per unit length $(l)$ for which we compute first $\mathbf{J}=J_z \mathbf{\hat{k}}$ from \eqref{eq:2dcurl} and evaluate the magnetization by
\begin{equation}
    \mathbf{M} = \frac{l}{2}
    \int_{\Omega_{\mathrm{sc}}} \mathbf{r}\times \mathbf{J}\dd \Omega.
    \label{eq:magnetization}
\end{equation}
Because this quantity is extracted from the same field representation used in the solve, it belongs naturally to the same residual-based workflow.


\begin{figure}[t]
    \centering
\includegraphics[width=0.85\textwidth]{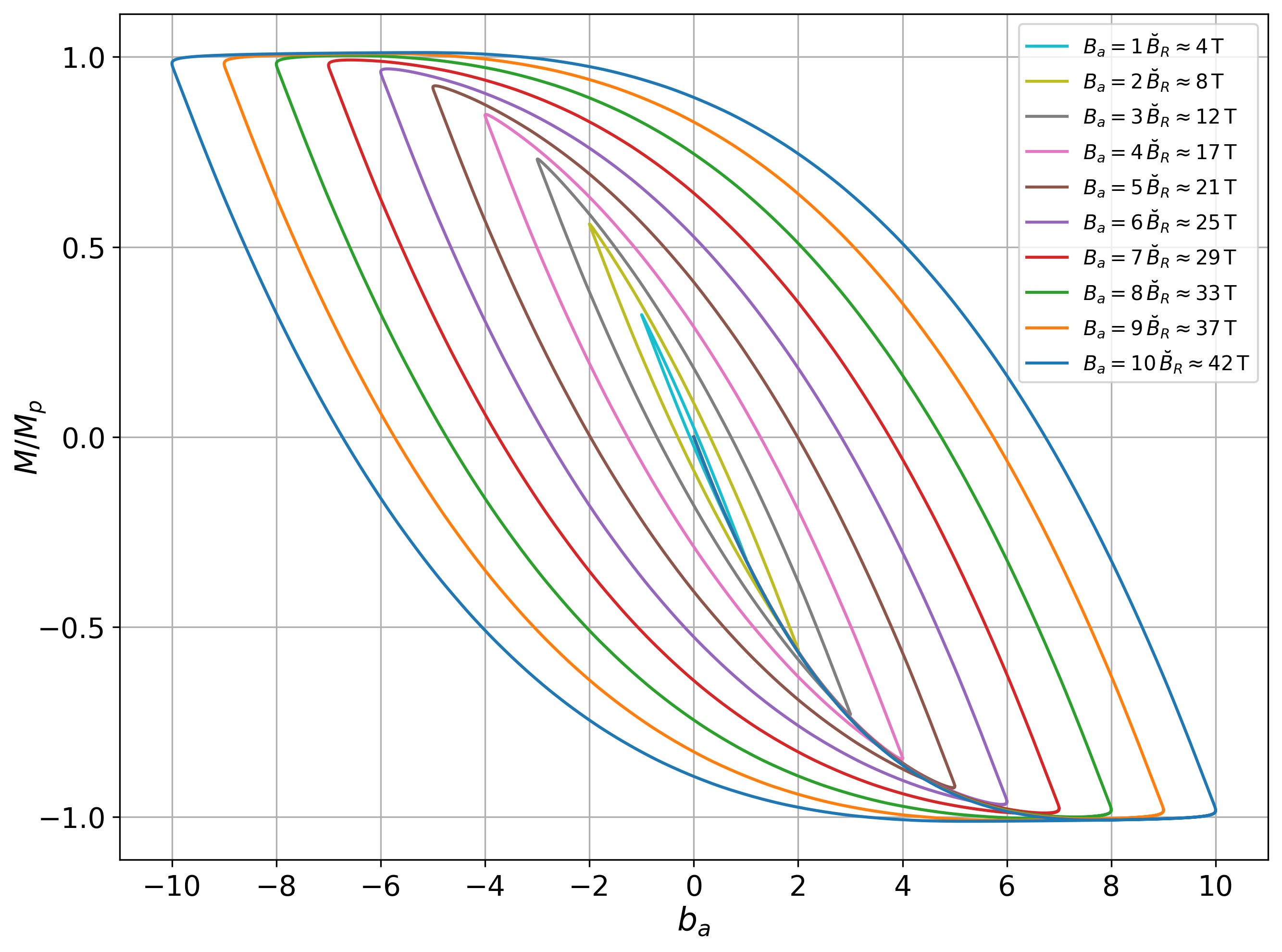}
    \caption{Calculated magnetization loops within the \emph{H-cloud} formalism for the cylindrical superconducting benchmark under AC transverse magnetic field of amplitude \(b_a=1,2,\ldots,10\) in units of $\check{B}_{R}=(\mu_0/4\pi)J_{c}R$. The magnetic moment is normalized by the analytical full-penetration value \(M_p\).}
    \label{fig:cylinder_verification}
\end{figure}


Analytically speaking, Bean's model allows to predict the full field penetration within the superconducting cylinder at $B_{p} = 8(\mu_0/4\pi)J_{c} R$, with the corresponding saturation of the magnetic moment achieved at $M_p = (2/3\pi)J_{c} R^{3}$. This corresponds to an state when the flux-free core \emph{disappears} satisfying the condition $0<|\mathbf{J}|\leq J_{c}$ under the $E-J$ power law, or strictly speaking $|\mathbf{J}|\equiv J_{c}$ under the pure critical state condition~\cite{Bean1964,Ruiz2012-APL}. This provides a direct test of whether the computed current penetration and magnetization response occur at the correct scale for any electromagnetic formulation in applied superconductivity. Yet, these exact solutions for $B_p$ and $M_p$ can be only obtained if the material law governing the superconducting state is unequivocally defined by the pure critical state condition for all $E \neq 0$ at any element within $\Omega_{\mathrm{sc}}$. However, due to its inequality character such material law cannot be easily incorporated in PDE commercial platforms, or at least not in a strict mathematical sense as it might be the case in integral-based formulations~\cite{Ruiz2010-SUST}. Therefore, in forward PDE formulations as the ones here discussed, the numerical model commonly uses a finite-\(n\) power-law \(E\)--\(J\) relation, as the one described in \eqref{eq:rho_piecewise}. Consequently, although the magnetization saturation will not be expected to resemble the characteristic flat plateau predicted by Bean at fields greater than $B_{p}$, the numerical solution must be as close as possible to it,  supported by a stringent sensitivity analysis. Thus, we use a slow excitation frequency $f=5\times10^{-4}\,\mathrm{Hz}$ and a high power-law exponent $n=96$, which places the solution close to the critical-state limit while retaining the same numerical formalism used elsewhere in the paper. 

Consistent with other established formulations and Bean’s analytical solution~\cite{Bean1964,Ruiz2012-PhDThesis}, \autoref{fig:cylinder_verification} shows the normalized magnetization loops obtained by increasing the maximum applied-field amplitude from \(b_a=1\) to \(10\), where $b_{a}=B_{a} / \check{B}_{R}$, with the renormalization field $\check{B}_{R}=(\mu_{0}/4\pi)J_{c}R$ which corresponds to exactly ${1}/{2\pi}$ times the magnetic field at the surface of a long, straight metallic (non-superconducting) wire of radius $R$ carrying a uniform current density \(J_{c}\). Note that, in these units, we can write \(B_p=8\check{B}_{R}\). To compare the two models, we compute
the superconductor magnetization for different incident field
intensities at their time of peak intensity $t^* = 3\pi/(2\omega)$. We denote this quantity by $M^*(B_a)$ and
compute the normalized magnetization increment
\begin{equation}
        \epsilon({b_a}, b_a+\Delta b_{a})\coloneqq\frac{|M^*(B_{a}+\check{B}_{R})-M^*(B_a)|}{M_p}\,.
    \label{eq:saturation_diagnostic}
\end{equation}

Considering a tolerance criterion of \(10^{-3}\), we say the superconductor is then fully penetrated at $b_a$ when the normalized magnetization increment
\eqref{eq:saturation_diagnostic} with $\Delta b_{a}=1$ falls first below
this tolerance.

The attained values are shown in \autoref{tab:plateau_check}, demonstrating how the resulting magnetization increment adequately falls below the prescribed tolerance once the applied field reaches the analytically predicted value of $b_a=8$, when any flux free core vanishes for fulfilling the critical threshold $|J_{z}|\simeq J_{c}$. This confirms therefore that the numerical solution by the \emph{H-cloud} model reproduces accurately the expected full-penetration scale within the benchmarked conditions. 

\begin{table}[t]
\centering
\caption{Magnetization Saturation Benchmarking.}
\label{tab:plateau_check}
\begin{tabular}{cccc}
\hline
$b_{a}$  & $\epsilon({b_a}, b_a+1)$ & State \\
\hline
6 & \(1.78\times10^{-2}\) & Unsaturated; Flux-free core clearly visible with $|J| = 0 \ | \sim 10\% \Omega_{sc}$\\
7 & \(3.05\times10^{-3}\) & Unsaturated; Flux-free core not fully evident, with $|J| \neq J_{c} \, \forall \, \Omega_{SC}$ \\
8 & \(6.46\times10^{-4}\) & Saturated; No flux-free core, $|J| \approx J_{c} \, \forall \, \Omega_{SC}$ \\
9 & \(5.55\times10^{-4}\) & Saturated; No flux-free core, $|J| \approxeq J_{c} \, \forall \, \Omega_{SC}$  \\
\hline
\end{tabular}
\end{table}

\begin{figure}[t]
    \centering
    \begin{subfigure}{0.31\textwidth}
        \centering
\includegraphics[width=\textwidth]{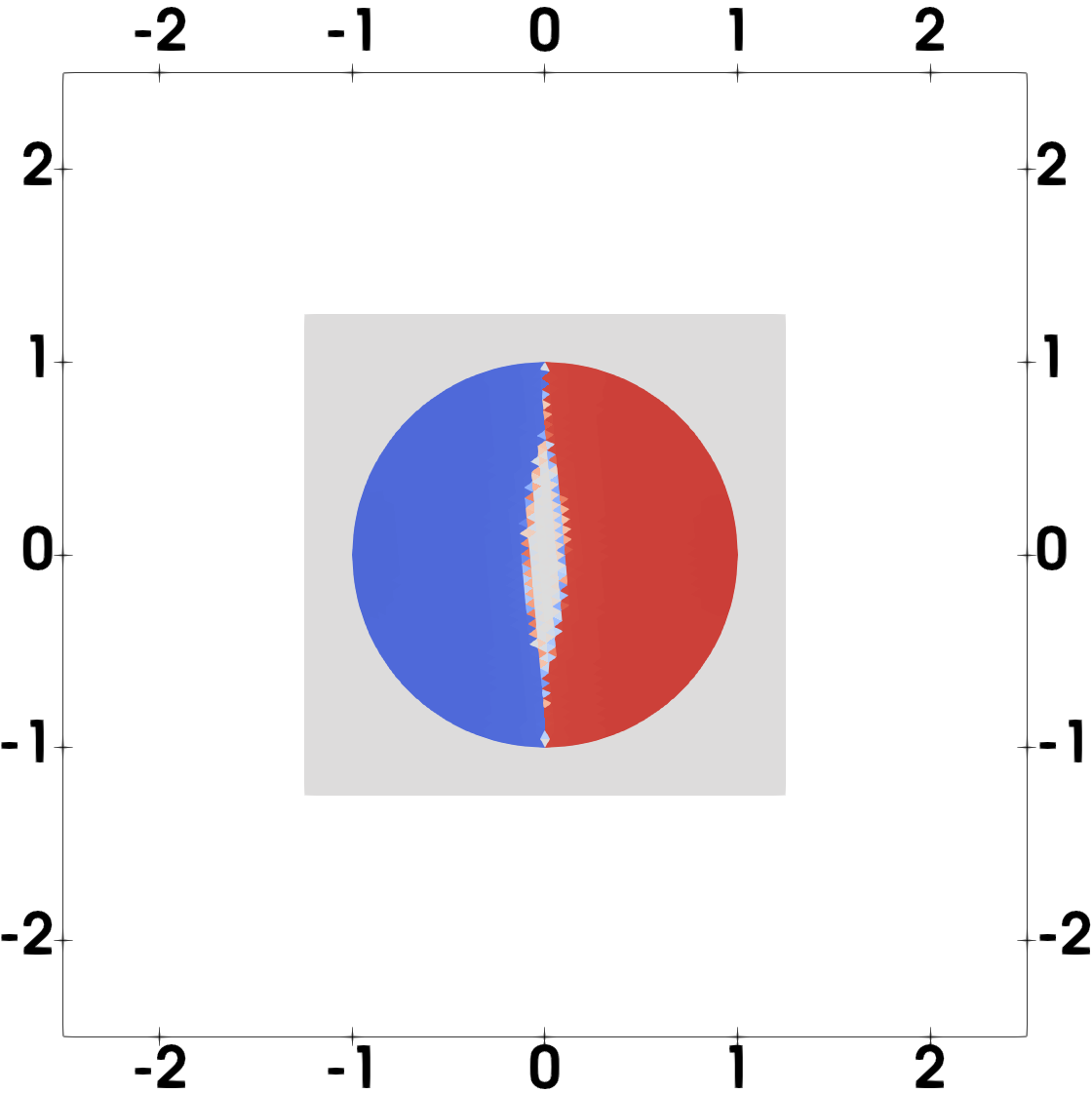}
        \caption{$max\left(|x|,|y|\right)= 1.25R$}
        \label{fig:figure2a}
    \end{subfigure}
    \hfill
    \begin{subfigure}{0.31\textwidth}
        \centering
\includegraphics[width=\textwidth]{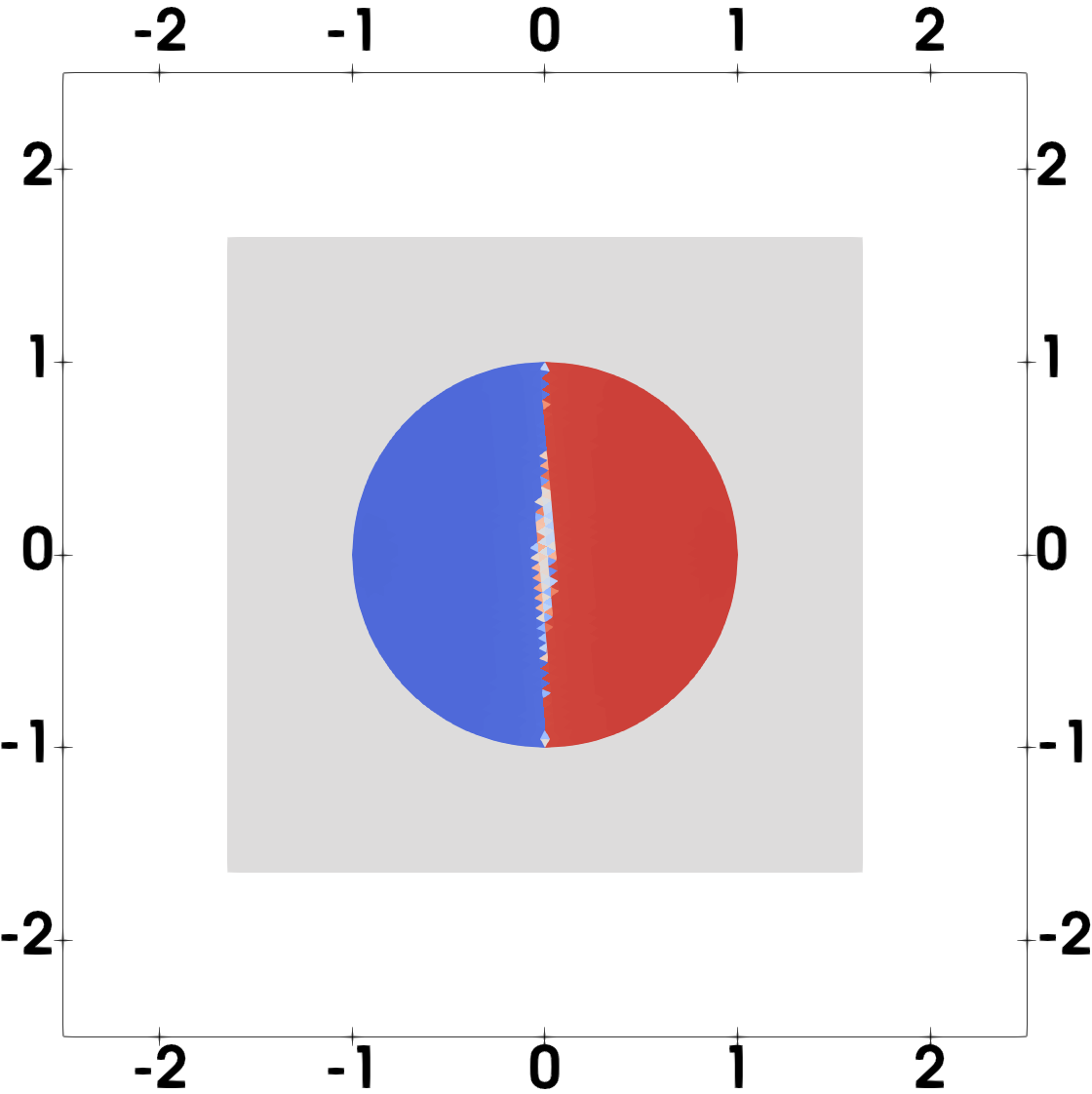}
        \caption{$max\left(|x|,|y|\right)= 1.65R$}
        \label{fig:figure2b}
    \end{subfigure}
    \hfill
    \begin{subfigure}{0.31\textwidth}
        \centering
\includegraphics[width=\textwidth]{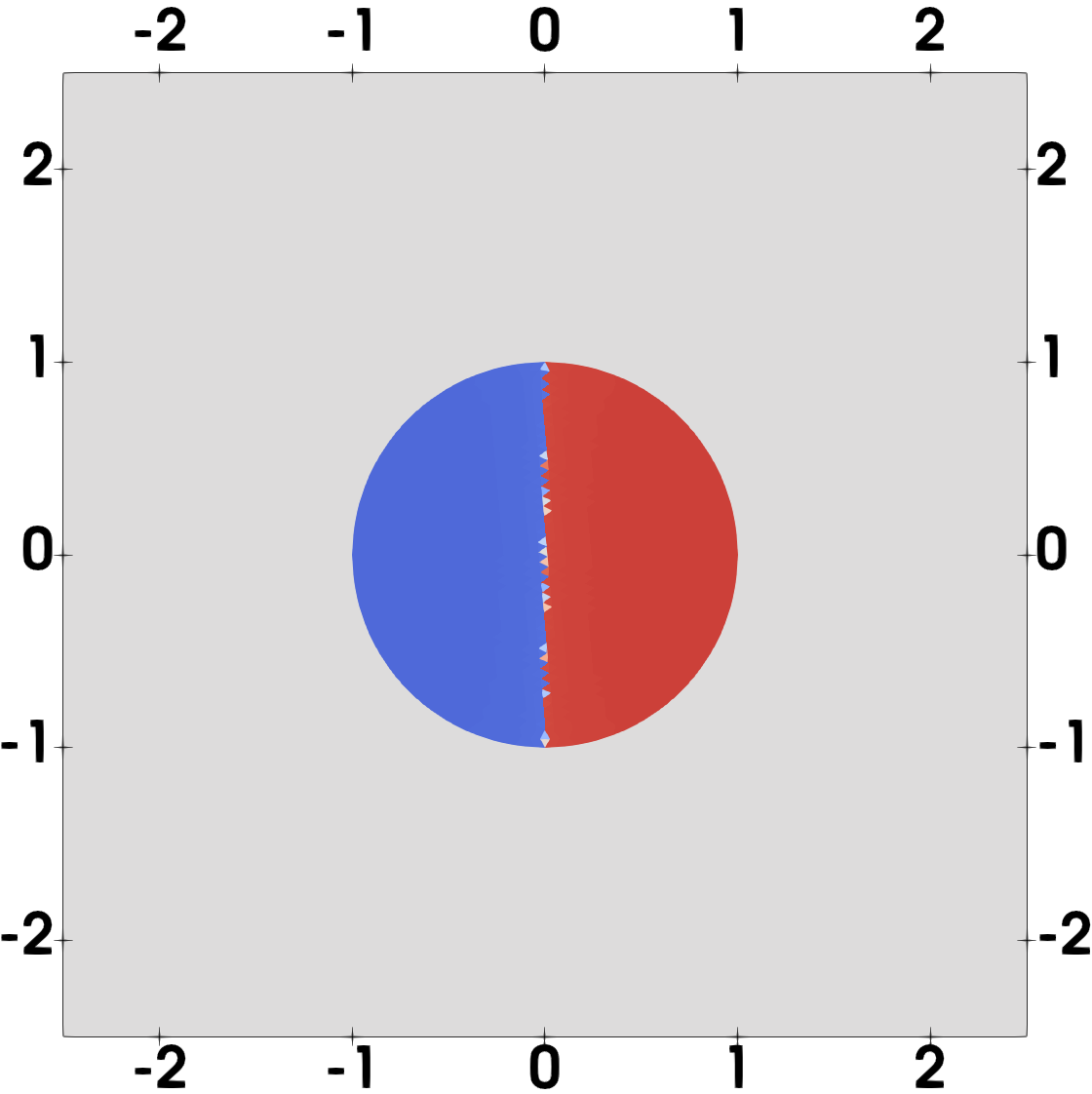}
        \caption{$max\left(|x|,|y|\right)= 2.5R$}
        \label{fig:figure2c}
    	\end{subfigure}
	\caption{Current density profiles at \(B_a=B_p=8\check{B}_{R}\) with the SC wire centred at $(0,0)$, for three surrounding boundary-domain sizes delimited by $\max(|x|,|y|)$ relative to (a) $1.25R$, (b) $1.65R$, and (c) $2.5R$. The comparison shows that a close outer boundary artificially suppresses flux penetration, whereas a sufficiently large air domain recovers the expected fully penetrated state at the expected analytical solution.}
    \label{fig:figure2}
\end{figure}

Similarly, \autoref{fig:figure2} also clarifies why a sensitivity analysis and benchmarking exercise constitute a useful test of the formulation itself. Since the applied field is imposed through \eqref{eq:tangential_bc}, a situation that arises in virtually any PDE formulation, whether field- or vector-potential-based, the finite truncation of the surrounding domain appears explicitly in the mathematics and should therefore be examined directly. For example, in COMSOL Multiphysics, arguably the most widely used finite-element PDE software for modelling superconducting systems, the applied-field condition may be imposed through high-level boundary-condition settings and thus appears less directly connected to the underlying weak formulation. Nevertheless, its mathematical treatment is ultimately the same. In other terms, while in the analytical benchmark the transverse field is prescribed at infinity, in a finite-element model it is imposed on the outer boundary of the computational domain. If this boundary is placed too close to the superconductor, the surrounding field is artificially constrained and full penetration is delayed. Enlarging the domain removes this artefact and restores the expected current distribution profile where no flux free core is expected at $B_a=B_p$ (see \autoref{fig:figure2}), fulfilling the benchmark conditions.

\subsection{Runtime benchmark}\vspace*{.2cm}

We next compare the proposed cloud-accessible workflow against an independently constructed COMSOL model for the same cylindrical benchmark in a practical high temperature superconductor setting. The case corresponds to a Bi2212-wire inspired-benchmark using representative material parameters, with $n=25$, $J_c= 6.64\times10^9\,\mathrm{A\,m^{-2}}$, $E_c=10^{-4}\,\mathrm{V\,m^{-1}}$, and $R=6.25~\mathrm{mm}$~\cite{Barua2021Bi2212,Oloye_2021,Jiang2019-Bi2212-TASC}. The outer boundary is placed at $10R$, and the transient solve spans $1.25$ cycles with $\Delta t=10^{-4}\,\mathrm{s}$, giving $250$ timesteps at $f=50\,\mathrm{Hz}$. In both cases we employ linear N\'ed\'elec
elements of the first kind, which are called ``linear curl elements'' in COMSOL.

The benchmark addresses two questions. First, does the open-source residual formulation reproduce the same magnetization process as a commercial finite-element implementation when geometry, material parameters, mesh scale, and time step are matched? Second, can the identical Python workflow be executed in a browser-based cloud environment while retaining the same numerical model? The primary metric is the normalized magnetization loop \(m=M/M_p\), since this quantity integrates the current-density distribution over the superconducting cross-section and is less sensitive to small pointwise differences between independent meshes. For each mesh level, the relative loop error $e_M$ is measured by the (discretised) $\mathrm{L}^2$ norm defined as
\begin{equation}
e_M =
\frac{
\left(\sum_k \Delta t |m_{\Hcurl}(t_k)-m_{\mathrm{COMSOL}}(t_k)|^2\right)^{1/2}
}{
\left(\sum_k \Delta t |m_{\mathrm{COMSOL}}(t_k)|^2\right)^{1/2}
} \, , 
\label{eq:loop_error}
\end{equation}

where $t_k$ is the time step, and the relative peak error is measured by the (discretized) $\mathrm{L}^\infty$ norm defined as,
\begin{equation}
e_{\mathrm{peak}} =
\frac{
\left|\max_k |m_{\Hcurl}(t_k)|-\max_k |m_{\mathrm{COMSOL}}(t_k)|\right|
}{
\max_k |m_{\mathrm{COMSOL}}(t_k)|
}.
\label{eq:peak_error}
\end{equation}

Worth mentioning, the values of $m_{\Hcurl}$ have been computed using a
fixed time step $\Delta t = 10^{-4}$ in \eqref{eq:be}. In this sense, athough
COMSOL uses an adaptive timestepping method, we retrieved the values of
$m_{\mathrm{COMSOL}}$ on the same $\Delta t$ time grid by specifying
the output times. \autoref{tab:runtime_accuracy_comparison} shows that the cloud and local open-source implementations of our $\Hcurl$-\textit{conforming} formulation, or \emph{H-cloud} for brief, reproduce the COMSOL magnetization loop with more than \(99\%\)-level of accuracy, while relative peak errors remain below \(1.5\%\). This difference is numerically negligible and can be reasonably attributed to the use of independently generated meshes, together with different nonlinear-solver implementations at each platform. More importantly, the cloud and local open-source runs produce essentially the same error levels, confirming that the hosted notebook execution changes the hardware environment but not the numerical formulation.

\begin{table}[t]
\centering
\caption{Runtime--accuracy comparison for the \emph{H-cloud} model operating within a cloud (Google Colab), or local environment, benchmarked against an equivalent case-model over matched mesh and timestep studies in a competitively commercial FEM-PDE platform (COMSOL Multiphysics 6.4).}
\label{tab:runtime_accuracy_comparison}
\begin{threeparttable}
\renewcommand{\arraystretch}{1.2}
\begin{tabular}{lllccc}
\hline
Platform & Hardware$^{*}$ & Mesh$^{\dag}$ & Runtime & \(e_M\) & \(e_{\mathrm{peak}}\) \\
\hline
Google Colab & Xeon, 2 logical cores & & 16 s & $7.75\times10^{-3}$ & $8.51\times10^{-3}$ \\
Firedrake Local & Apple M3, 8 CPU cores & coarse & 12 s & \(7.75\times10^{-3}\) & \(8.56\times10^{-3}\) \\
COMSOL Local & i5-12400, 6 CPU cores &  & 13 s & reference & reference \\
\hline
Google Colab & Xeon, 2 logical cores &  & 189 s & \(1.304\times10^{-2}\) & \(1.372\times10^{-2}\) \\
Firedrake Local & Apple M3, 8 CPU cores & medium & 45 s & \(1.304\times10^{-2}\) & \(1.372\times10^{-2}\) \\
COMSOL Local & i5-12400, 6 CPU cores &  & 57 s & reference & reference \\
\hline
Google Colab & Xeon, 2 logical cores &  & 1752 s & \(1.427\times10^{-2}\) & \(1.483\times10^{-2}\) \\
Firedrake Local & Apple M3, 8 CPU cores & fine & 407 s & \(1.428\times10^{-2}\) & \(1.455\times10^{-2}\) \\
COMSOL Local & i5-12400, 6 CPU cores &  & 392 s & reference & reference \\
\hline
\end{tabular}
\renewcommand{\arraystretch}{1.0}
\begin{tablenotes}
       \item $^{*}$ \footnotesize{Hardware is reported to contextualize runtime differences, not as a processor-normalized performance study.}
       \item $^{\dag}$ \footnotesize{The finite elements number at coarse, medium, and fine meshes are 1916, 7664, and 30656 elements respectively.}
\end{tablenotes}
\end{threeparttable}
\end{table}

The local runtime comparison shows that the \emph{H-cloud} implementation in Firedrake and the COMSOL Multiphysics solution both operate at the same order of computational cost for this benchmark. However, the cloud results must be interpreted more carefully because their hardware specs are not equivalent. The \emph{H-cloud} local simulations were run on an 8-core Apple M3 MacBook with 16 GB RAM. The COMSOL simulations were run on a Windows 11 PC with an Intel Core i5-12400 desktop processor, six CPU cores, a 2.5 GHz base clock, and 15.92 GB RAM. By contrast, the Google Colab run used a representative free CPU session with an Intel Xeon processor at \(2.20\,\mathrm{GHz}\), two logical CPU cores, and 12.67 GB RAM. Thus, the local machine had approximately four times the number of CPU cores available to the Colab session and a substantially higher performance-core clock speed. The medium and fine cloud runtimes, which are approximately four times longer than the local runtimes, are therefore consistent with the reduced CPU allocation rather than with a change in the numerical model. A local Linux implementation of the same \emph{H-cloud} workflow was also tested, rendering to essentially the same computation time as the macOS implementation.

The central computational result is therefore not raw cloud-runtime superiority, but the portability and reproducibility of the \emph{H-cloud} method across local and cloud-executable environments. The same weak form, mesh hierarchy, material law, boundary condition, solver configuration, and post-processing can be executed in Google Colab without a local installation or commercial licence. The measured peak process memory was only 0.46 GB for the medium mesh and 0.51 GB for the fine mesh, indicating that the present examples are CPU-time limited rather than memory limited on Colab. For larger problems, the same notebook-based workflow could be moved to stronger cloud resources, including paid Colab runtimes, Kaggle notebooks, Amazon SageMaker Studio or Studio Lab, Paperspace Gradient, or dedicated cloud instances. Moreover, because the implementation is built on automated finite-element code generation and PETSc solver infrastructure, future acceleration routes include multi-core CPU deployments and, subject to backend support for the chosen finite-element kernels and solvers, GPU-enabled cloud or high-performance-computing environments. This provides a practical route to open, reproducible superconducting simulations that can be accessed by researchers without specialised local infrastructure, while remaining extensible to more demanding computational resources and environments where portability and compatibility of non-black-box constructions for computations is deemed essential for scalability.


\section{Conclusion} 

In this paper, we have introduced \emph{H-cloud}, a variationally transparent formulation of the superconducting magneto-quasi-steady problem implemented as a cloud-accessible, open-source and reproducible Python-based finite-element framework. Built upon the most general of the PDE formulations in applied superconductivity, the so-called $H$-formulation, the method recasts the superconducting electromagnetic problem into an explicit curl-conforming discretisation, thereby linking the mathematical model directly to an executable and modifiable computational workflow. By expressing the governing problem in the curl-conforming function space 
\(H(\mathrm{curl})\), with the applied-field tangential trace, nonlinear material law, finite-element discretisation, and implicit time stepping all stated at the variational level, \emph{H-cloud} preserves the governing physics of the classical $H$-formulation while exposing the full computational pathway for portability and reproducibility in physically local and cloud architectures. Moreover, the same residual also provides a natural entry point for PDE-constrained optimisation, connecting forward superconducting simulation to adjoint-based design and inverse problems within a unified framework.

The numerical results shown that this formulation is both physically reliable and computationally portable. The Bean-cylinder magnetisation benchmark recovers the expected full-penetration scale, confirming that the method resolves the critical-state behaviour that underpins canonical superconducting validation tests. In addition, the runtime--accuracy comparison against an independently constructed COMSOL model showed that the local and cloud-based open-source implementations reproduce the same magnetisation response to within approximately \(1\%\) in loop error, while relative peak errors remain below \(1.5\%\) across the tested mesh hierarchy. The cloud execution therefore changes the hardware context, but not the mathematical model or the numerical outcome. In fact, rather than treating cloud execution as a secondary deployment choice, the present work shows that portability can emerge naturally from a well-posed variational construction.

Taken together, these results establish the \emph{H-cloud} approach as a novel, rigorous, and portable route for superconducting electromagnetic modelling, which can be formulated, shared, and executed as a transparent scientific object rather than as a computational platform-dependent recipe. Yet, it is not intended to replace nor compete with any commercial PDE software, nor to favour a formalism choice for the electromagnetic modelling of superconductors, as such choice is case-based. Instead, it aims to broaden access to advanced modelling tools in applied superconductivity, provide a practical foundation for extensive benchmark sharing, collaborative model development and even teaching, without requiring proprietary licences or extensive local setup (see \emph{Supplementary Material}). Additionally, the framework established here provides a strong basis for future extensions to more complex geometries, richer constitutive laws, inverse problems~\cite{Ruiz2025-SciAdv}, and scalable cloud or high-performance-computing deployments. This gives \emph{H-cloud} a unique methodological flexibility, both mathematical and computational, allowing it to extend coherently from reproducible benchmark studies towards more demanding superconducting device simulations and optimisation-led design workflows.

\ack{}
This work was supported in part by the U.K. Research and Innovation (UKRI), Engineering and Physical Sciences Research Council (EPSRC), under grant Ref. EP/S025707/1 led by H.S.R. The work of Y.G. was supported by the Future 100 PhD programme of the University of Leicester.

\roles{Conceptualization: Y.G., A.P., H.S.R;
Methodology: Y.G., A.P., H.S.R; Investigation: Y.G., A.P., H.S.R; Visualization: Y.G., H.S.R.; Results Analysis: A.P., H.S.R.; Funding acquisition: A.P., H.S.R.; Project administration: A.P., H.S.R.; Supervision: A.P., H.S.R.; Writing – original draft: Y.G.; Writing – review \& editing: A.P., H.S.R.}

\section*{Additional information}
\textbf{Competing interests:} The authors have no competing interests to declare. 

\noindent\textbf{Data and materials availability:} All data needed to evaluate and reproduce the Results and Discussion in the paper are present in the paper and/or the \emph{Supplementary Material}. Source data and code are also available in DOI: 10.5281/zenodo.21294926.

\noindent\textbf{License information:} For the purpose of open access, the authors have applied a Creative Commons Attribution (CC BY) licence to the Author Accepted Manuscript version arising from this submission.

\bibliographystyle{iopart-num}
\bibliography{Main_References_June_2026}
\end{document}